\documentclass{emulateapj}
\submitted{{\it accepted for publication in AJ}}
\usepackage{graphicx}
\shortauthors{West et al.}
\shorttitle{DR5 Low-Mass Stars}
\begin{document}

\title{Constraining the Age-Activity Relation for Cool Stars: The SDSS
  DR5 Low-Mass Star Spectroscopic Sample}

\author{Andrew A. West\altaffilmark{1,2}, 
Suzanne L. Hawley\altaffilmark{3},
John J. Bochanski\altaffilmark{3},
Kevin R. Covey\altaffilmark{4,5},
I. Neill Reid\altaffilmark{6},
Saurav Dhital\altaffilmark{7},
Eric J. Hilton\altaffilmark{3},
Michael Masuda\altaffilmark{3}}

\altaffiltext{1}{Corresponding author: awest@astro.berkeley.edu}
\altaffiltext{2}{Astronomy Department, University of California, 601 Campbell Hall, Berkeley, CA 94720-3411}
\altaffiltext{3}{Department of Astronomy, University of Washington, Box 351580,
Seattle, WA 98195}
\altaffiltext{4}{Harvard-Smithsonian Center for Astrophysics, 60
  Garden Street, Cambridge MA 02138}
\altaffiltext{5}{Spitzer Fellow}
\altaffiltext{6}{Space Telescope Science Institute, 3700 San Martin Drive,Baltimore, MD 21218}
\altaffiltext{7}{Department of Physics \& Astronomy, Vanderbilt
  University, Nashville, TN 37235}

\begin{abstract} 
We present a spectroscopic analysis of over 38,000 low-mass stars from
the Sloan Digital Sky Survey (SDSS) Data Release 5 (DR5). Analysis of
this unprecedentedly large sample confirms the previously detected
decrease in the fraction of magnetically active stars (as traced by
H$\alpha$ emission) as a function of vertical distance from the
Galactic Plane.   The magnitude and slope of this effect varies as a
function of spectral type.  Using simple 1-D dynamical models, we
demonstrate that the drop in activity fraction can be explained by
thin disk dynamical heating and a rapid decrease in magnetic
activity.  The timescale for this rapid activity decrease changes
according to the spectral type. By comparing our data to the
simulations, we calibrate the age-activity relation at each M dwarf
spectral type.  We also present evidence for a possible decrease in
the metallicity as a function of height above the Galactic
Plane.  In addition to our activity analysis, we provide line
measurements, molecular band indices, colors, radial velocities, 3-D
space motions and mean properties as a function of spectral type for
the SDSS DR5 low-mass star sample.
\end{abstract}


\section{Introduction}

As the most numerous stellar constituents of the Milky Way, M dwarfs
are an ideal population for tracing the structure and evolution of the
stellar thin disk.  A number of studies have utilized the ubiquity of
M dwarfs to study the dynamics and distribution of stars in the Solar
neighborhood (Wielen 1977; Weis \& Upgren 1995; Reid et al. 1995;
Hawley et al. 1996; Bochanski et al. 2007b).  Many M dwarfs also
host intense magnetic dynamos that give rise to chromospheric and
coronal heating, producing emission from the x-ray to the radio.  Over
30 years ago Wilson and Woolley (1970) found a link between
magnetic activity (as traced by the CaII emission strength) and the
orbits of more than 300 nearby late-type dwarfs.  Their study
concluded that magnetic activity in these stars was directly related to
their age.  Subsequent studies over the following decades have found a
similar connection between age and activity in low-mass stars (Wielen
1977; Giampapa \& Liebert 1986; Soderblom, Duncan \& Johnson 1991;
Hawley et al. 1996; Hawley, Tourtellot \& Reid 1999; Hawley, Reid \&
Tourtellot 2000). 

The mechanism that controls magnetic activity in M dwarfs is still
unknown.  In the Sun, magnetic field generation has a strong
connection to the Sun's rotation.  Helioseismology has indicated that
the boundary between the convective and the radiative zones (known as
the tachocline) creates a rotational sheer; the convective zone
undergoes differential rotation, while the radiative zone rotates like
a solid body (Parker 1993; Ossendrijver 2003; Thompson et al. 2003).
The tachocline allows magnetic fields to be generated, stored and
ultimately rise to the surface where they emerge as magnetic loops.
These loops drive the heating of the stellar chromosphere and corona,
resulting in dramatic stellar flares and lower level quiescent magnetic activity.  Rotation in solar type stars slows with
time and as a result, activity decreases.  Skumanich (1972) found that
both activity (as measured by Ca II emission) and rotation decrease
over time as a power law t$^{-0.5}$.  Subsequent studies confirmed the
Skumanich age/rotation-activity relation but found that the slope of
the power law may change as a function of spectral type and that
non-power law functions are not ruled out (Barry 1988; Soderblom et
al. 1991). However, after a spectral type of $\sim$M3 (0.35
M$_{\odot}$; Reid \& Hawley 2005; Chabrier \& Baraffe 1997), stars become fully
convective and the tachocline presumably disappears. This transition marks an important change in the
stellar interior that must affect the production and storage of
internal magnetic fields.


Many studies have found evidence that the age-activity relation
extends into the M dwarf regime.  Eggen (1990) observed a Skumanich
type power law decay in activity strength as a function of age.
Larger samples of M dwarfs have added further evidence to the
hypothesis that the magnetic activity in M dwarfs slowly decreases
over time (Fleming, Schmitt \& Giampapa 1995; Gizis, Reid \& Hawley
2002).  There are also data that support the hypothesis that activity
in M dwarfs may have a finite lifetime.  Stauffer et al. (1994)
suggested that activity may not be present in the more massive M
dwarfs in the Pleiades.  Hawley et al. (2000) confirmed the Stauffer
et al. (1994) claim by observing a sample of clusters that spanned
several Gyrs in age.  They were able to calibrate the activity
lifetimes for early type M dwarfs by observing the color at which
activity (as traced by spectroscopically observed H$\alpha$ emission)
was no longer present.  Because of the small sample size, the derived
age-activity relation serves as a lower limit at a given color or
spectral type.  Silvestri et al. (2006) also argued for finite
activity lifetimes by observing M dwarfs in binary systems with white
dwarfs, finding no activity in the companions of the coolest, and
therefore oldest white dwarfs.  However, accurate activity lifetimes
could not be calculated for the M dwarfs due to uncertainties in the
pre-white dwarf ages, binary effects and small sample sizes.  A larger
spectroscopic sample is required to further constrain the activity-age
relation in both nearby clusters and the field.

The advent of large surveys such as the Sloan Digital Sky Survey
(SDSS; York et al. 2000) and the Two Micron All Sky Survey (2MASS;
Cutri et al. 2003) has created optical and infrared catalogs of
several million M dwarfs.  In addition to the photometric data, the
SDSS has obtained spectra for over 50,000 M dwarfs and the numbers
continue to increase as SDSS-II (Adelman-McCarthy et al. 2007b) pushes
to lower Galactic latitudes.  Recently these surveys have been
utilized to examine the statistical properties of M dwarfs in a
variety of capacities, including analyses of mean M dwarf colors and
spectroscopic properties (Hawley et al. 2002; West et al. 2004;
hereafter W04; West, Walkowicz \& Hawley 2005; Bochanski et al. 2007a;
Covey et al. 2007), computing the dynamics of the Milky Way (West et
al. 2006; hereafter W06; Bochanski et al. 2007b) and the detailed
investigation of chromospheric magnetic activity (W04; W06; Silvestri
et al. 2006; Bochanski et al. 2007a).  The activity studies of W04 and
W06 used SDSS spectra to show that the fraction of active M7 stars
decreases with vertical distance from the Galactic Plane, further
evidence of an age-activity relation. By comparing the data to simple
dynamical simulations, W06 argued that the observed trend results from
the combination of a rapid decrease in magnetic activity after 6-7
Gyrs and the correlation between age and height above the Galactic
Plane introduced by dynamical heating of the M7 stars. A larger sample
of low-mass star spectra allows for the obvious extension of the W06
study to all M dwarf spectral types.  From this new statistical
platform it is now possible to derive an age-activity relation for the
entire M dwarf sequence and to investigate how this relation changes
across the M3-M4 convective boundary.

In this paper we present a catalog of more than 44,000 spectroscopically
confirmed M dwarfs from the SDSS.  We describe the data, which are
publically available, in \S2.  We present the mean sample characteristics and
an analysis of the magnetic activity in \S3. In \S4, we compare our data to 
a series of dynamical simulations and derive an age-activity relation
for low-mass dwarfs.  We summarize these results in \S5.

\section{Data}

\begin{figure}
\plotone{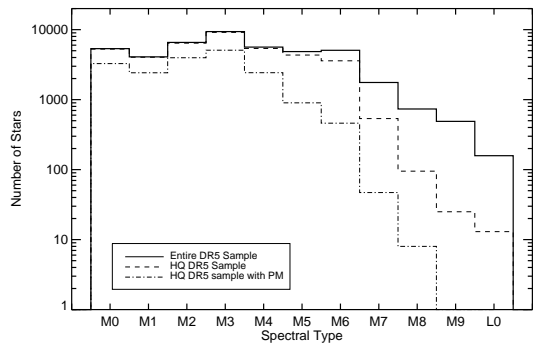}
\caption{Spectral type distribution of the SDSS DR5 low-mass star
  spectroscopic sample.  The total sample (solid) contains over 44,000
  spectroscopically confirmed M and L-type dwarfs. Our
  analysis is confined to a high quality subsample (dashed) and a high quality
  subsample with measured proper motions (dot-dashed) which have over
  38,000 and 20,000 stars respectively.  See text for more details.}
\label{figure:dr5dist} 
\end{figure}

The SDSS (Gunn et al. 1998; Fukugita et al. 1996; York et al. 2000;
Hogg et al. 2001; Gunn et al. 2006; Ivezi{\'c} et al. 2004; Pier et
al. 2003; Smith et al. 2002; Stoughton et al. 2002; Tucker et al. 2006) provides excellent
quality spectroscopic and photometric data for examining the
properties of low-mass dwarfs. Our sample was drawn from the SDSS Data
Release 5 (DR5; Adelman-McCarthy 2007a) and includes spectra from
several of the ``special plates'' that are not part of the standard
SDSS survey\footnote{see
  http://www.sdss.org/dr5/products/spectra/special.html}.  A total of
49933 stars were selected from the SDSS spectroscopic database based
on the colors typical of M dwarfs ($r-i >$ 0.53 and $i-z >$ 0.3; West
et al. 2005) and a standard set of SDSS processing flags (SATURATED,
BRIGHT, NODEBLEND, INTERP\_CENTER, BAD\_COUNTS\_ERROR, PEAKCENTER,
NOTCHECKED, and NOPROFILE were all required to be unset).\footnote{See
  http://www.sdss.org/dr5/products/catalogs/flags.html for more
  information on SDSS flags}  Each spectrum was processed using the
HAMMER stellar spectral-typing facility (Covey et al. 2007).  The
HAMMER determines the spectral type of each star using measurements of
molecular bands and line strengths.  We examined a subset of
the spectra by eye and confirmed that the spectral types given by the
HAMMER were accurate to within 1 subtype.  We also limited our study
to spectral types M0-L0.  The HAMMER was
able to successfully spectral type 44084 of the stars. Most of the
spectra that were not typed were either low signal-to-noise or bad
spectra.  The spectral
type distribution of the DR5 low-mass star 
spectroscopic sample is shown in Figure
\ref{figure:dr5dist}.

To ensure a statistically robust sample, we removed stars with
1) line-of-sight $r$-band extinction $>$ 0.5; 2) colors consistent with
having a white dwarf companion (Smol{\v c}i{\'c} et al. 2004); and 3) measured radial
velocities $>$ 500 km s$^{-1}$ (spurious velocities).  The resulting high quality dataset
contains 38835 stars, roughly 4 times the number of stars used in the
W04 study.

The magnetic activity of each star was assigned based on its measured
H$\alpha$ emission line strength.  Although other emission lines can
be used as activity indicators (e.g. CaII H \& K), our study defines
an active star as one that has H$\alpha$ emission in its spectrum.  H$\alpha$
EWs were measured by the HAMMER, which now includes most of the
magnetic activity analysis described in W04.  We assessed the accuracy
of our activity analysis by running Monte Carlo simulations to test
how well we could measure the H$\alpha$ emission line at a specific
spectral type and for a given signal-to-noise spectrum.  Using the
Bochanski inactive templates (Bochanski et al. 2007a), we added noise
and an emission line to each spectral type and then used our software
to measure the H$\alpha$ EW 1000 times for each variation of noise,
spectral type and emission strength. We found that at a
signal-to-noise ratio (measured near H$\alpha$) of 3 and an H$\alpha$
EW of 1\AA\ (the same parameters used in previous studies; e.g. W04,
W06), we could accurately recover the H$\alpha$ emission over 96\% of
the time for all spectral types.  This agrees with the less than 4\%
error in assigning activity status that was estimated in W04.

  After running the stars through the HAMMER, we
used the method of Walkowicz, Hawley \& West (2004) to compute the
ratio of luminosity in the H$\alpha$ line to the bolometric luminosity
(L$_{\rm{H\alpha}}$/L$_{bol}$) for all stars in the sample.  We
adopted the color derived ($i-z$) $\chi$ values for this
calculation. Distances to all stars were calculated using the
photometric parallax methods of West et al. (2005) and Davenport et
al. (2006). Using the positions and distances of each star, Galactic
height was computed assuming the Sun is 15 pc above the Plane (Cohen
1995; Ng et al. 1997; Binney et al. 1997).

\begin{deluxetable*}{lrrccccc}
\tablewidth{0pt}
\tablewidth{0pt}
\tablecolumns{8} 
\tabletypesize{\scriptsize}
\tablecaption{Colors of Late-Type Dwarfs}
\renewcommand{\arraystretch}{.6}
\tablehead{
\colhead{Spectral Type}&
\colhead{N$_{\rm{SDSS}}$\tablenotemark{a}}&
\colhead{N$_{\rm{2MASS}}$\tablenotemark{a}}&
\colhead{$r-i$}&
\colhead{$i-z$}&
\colhead{$z-J$}&
\colhead{$J-H$}&
\colhead{$H-K$}}
\startdata
M0 & 4945 & 1962 & 0.66 (0.12) &0.38 (0.07) &1.23 (0.19) &0.65 (0.10) &0.18 (0.14)\\
M1 & 3721 &1504 &0.82 (0.18) &0.46 (0.07) &1.27 (0.21) &0.64 (0.10) &0.21 (0.13)\\
M2 & 6105 & 2472 &1.00 (0.13) &0.55 (0.08) &1.31 (0.17) &0.62 (0.10) &0.22 (0.12)\\
M3 & 8609 & 3752& 1.21 (0.16) &0.65 (0.08) &1.37 (0.17) &0.60 (0.10) &0.24 (0.14)\\
M4 & 4984 & 2126&1.46 (0.15) &0.79 (0.09) &1.45 (0.16) &0.59 (0.10) &0.27 (0.13)\\
M5 & 3608 & 879& 1.91 (0.13) &1.06 (0.07) &1.61 (0.11) &0.59 (0.11) &0.31 (0.14)\\
M6 & 2674 & 570&2.11 (0.14) &1.16 (0.07) &1.71 (0.10) &0.60 (0.11) &0.33 (0.14)\\
M7 & 382 & 306&2.50 (0.18) &1.37 (0.09) &1.85 (0.10) &0.62 (0.09) &0.36 (0.11)\\
M8 & 74 & 147&2.73 (0.25) &1.58 (0.15) &2.04 (0.12) &0.66 (0.06) &0.40 (0.06)\\
M9 & 13 & 69&2.81 (0.21) &1.76 (0.08) &2.25 (0.09) &0.70 (0.08) &0.42 (0.06)\\
L0 & 10 & 31&2.49 (0.09) &1.84 (0.06) &2.45 (0.11) &0.77 (0.06) &0.50 (0.04)\\
\enddata
\tablecomments{The median and rms scatter (in parentheses) of the
  SDSS-2MASS colors at each spectral type.  Because the stars used to
  compute these values have small measurement errors, the scatter
  represents the intrinsic scatter of the stellar locus and not the
  uncertainties in the median measurement. SDSS spectroscopic
  targeting does not uniformly sample the stellar locus, possibly
  introducing a bias to our derived colors. The $u-g$ and $g-r$ colors
  are not shown because of the large influence that metallicity has on
  the $g-r$ colors (West et al. 2004; Bochanski et al. 2007a) and the
  large uncertainties in the $u$-band fluxes for most of the sample
  stars.}  \tablenotetext{a}{Only stars with magnitude uncertainties
  smaller than 0.05 were used to compute the median colors}
\label{table:colors}
\end{deluxetable*}

We used the SDSS/USNO-B matched catalog (Munn et al. 2004) to obtain
proper motions. The SDSS/USNO-B catalog uses SDSS galaxies to recalibrate
the USNO-B positions as well as includes SDSS stellar astrometry as an
additional epoch for improved proper motion measurements.  The
USNO-B proper motions were obtained for 27285 of the SDSS stars (many
stars were not matched due to the relative shallowness of the USNO-B catalog).
Applying the same quality cuts as above, we arrived at a proper motion
sample of 20160 stars.  Radial velocities were measured for all SDSS
stars using the method described in Bochanski et al. (2007a),
employing co-added, zero-velocity SDSS spectra which were used to
create high signal-to-noise ratio radial velocity templates.
Our derived radial velocities are accurate to within $\sim$5 km
s$^{-1}$ (Bochanski et al. 2007a).  Using these proper motions the
derived distances and the radial velocities, space
motions were computed for all SDSS stars with measured proper motions.

We also matched our spectroscopic sample to the 2MASS point source
catalog and found matches for 37845 of our stars.  Although the
infrared colors are not a vital part of our analysis, we include them
in order to compute statistically robust 2MASS colors as a function of
spectral type and will include them in the online electronic tables (see below).

\begin{figure}
\plotone{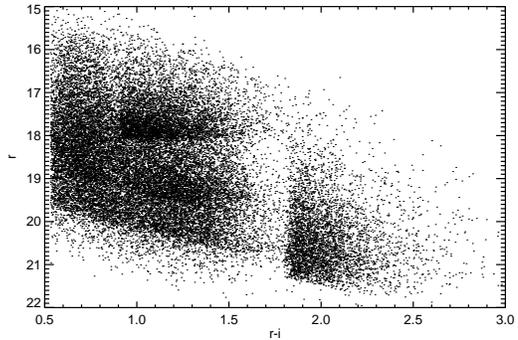}
\caption{The color-apparent magnitude distribution for the sample.
  The blue ``edge'' is our color selection described above.  The range
  of magnitudes is constrained by both the bright and faint limits of
  SDSS spectroscopy (Stoughton et al. 2002; Strauss et al. 2002) as well as our requirement
  that none of the stars be saturated in the photometric data.  The
  gaps and overdense regions in color-magnitude space do not represent
  the actual distribution of stars but are a reflection of the SDSS
  spectroscopic targeting selection criteria, which are not designed
  to sample stars in a complete, or even uniform, manner.}
\label{figure:cmd}
\end{figure}

The measured quantities for the SDSS DR5 low-mass star spectroscopic
sample are available for electronic download for use by the
astronomical community\footnote{Measured quantities can be obtained
  electronically using the CDS Vizier database
  http://vizier.u-strasbg.fr/viz-bin/VizieR}.  The individual spectra
are available at the SDSS catalog archive
server\footnote{http://cas.sdss.org/astrodr5/en}.  It is important to
note that these data do not represent a complete sample and that
spectral targeting introduces a variety of selection effects.
However, the sample covers a large range of values for many of the
physical attributes of the stars, including activity, metallicity and
Galactic motion, making an accurate activity analysis with this sample
possible.  In addition, because most of the derived quantities are
computed by automatic routines, values for a small percentage of
individual stars will be incorrect.  Large statistical results should
not be affected by these small errors.  Users are nevertheless
cautioned to understand the origins of the data before using them
indiscriminately.

\section{Observational Results}

Table \ref{table:colors} gives the median and rms scatter (in
parentheses) of the SDSS and 2MASS colors at each spectral type. The
stars used to compute these values were selected to have small
measurement errors and thus, the scatter represents the intrinsic
scatter of the stellar locus and not the uncertainties in the median
measurement.  SDSS spectroscopic targeting does not uniformly sample
the stellar locus, possibly introducing a bias to our derived colors.
The $u-g$ and $g-r$ colors were not shown because of the
large influence that metallicity has on the $g-r$ colors (West et
al. 2004; Bochanski et al. 2007a) and the large uncertainties in the
$u$-band fluxes for most of the sample stars.  The SDSS colors are in
excellent agreement with the previous results found in West et
al. (2005) and Bochanski et al. (2007a).  Because of the larger sample
size, our 2MASS colors supersede those reported in Hawley et
al. (2002), West et al. (2005) and Covey et al. (2007). Figure
\ref{figure:cmd} shows the color-apparent magnitude distribution for
 the sample.  The blue ``edge'' is our color selection described
 above.  The range of magnitudes is constrained by both the bright and
 faint limits of SDSS spectroscopy (Stoughton et al. 2002; Strauss et al. 2002) as well as
 our requirement that none of the stars be saturated in the photometric
 data.  The gaps and overdense regions in color-magnitude space
 do not
 represent the actual distribution of stars but are a reflection of
 the SDSS spectroscopic targeting selection criteria, which are
 not designed to sample stars in a complete, or even uniform, manner.
 Figure \ref{figure:cmd} also helps to demonstrate why some of the
 computed colors from Table \ref{table:colors} may be affected by selection biases.
 Clearly, an analysis of the luminosity function of low mass stars cannot
 be carried out with these data, but is currently being 
 addressed using the SDSS photometric data (Covey et
 al. 2008; Bochanski et al. 2008).

\subsection{Activity}

\begin{figure}
\plotone{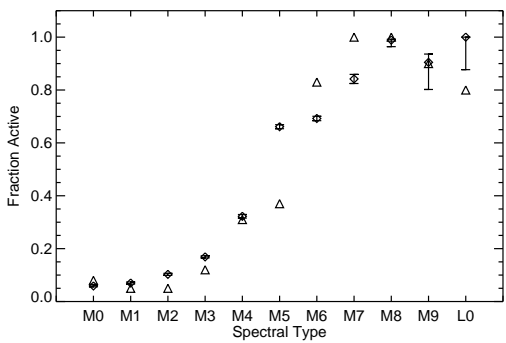}
\caption{Fraction of active stars as a function of spectral type for
  the DR5 spectroscopic sample (diamonds). At earlier types ($<$M7)
  this distribution resembles that of previous studies (triangles;
  Hawley et al. 1996; Gizis et al. 2000; W04).  However, we do not see
  a turnover in the activity fraction in the latest types as seen
  previously.  This is likely a result of our strict quality cuts,
  which remove the distant, late-type stars from the sample.  The more
  distant stars are more likely to be older and therefore
  inactive. However, the low signal-to-noise in those spectra does not
  allow us to be certain of their activity state and they are
  therefore excluded from the analysis.}
\label{figure:actfrac} 
\end{figure}

Applying the aforementioned signal-to-noise and EW cuts (3 and 1\AA\
respectively) to our activity analysis yields an updated distribution
of activity fraction as a function of spectral type (Figure
\ref{figure:actfrac}).  At earlier types ($<$M7) this distribution
resembles that of previous studies (Hawley et al. 1996; Gizis et
al. 2000; W04).  However, the later types do not show the same
decrease in activity fraction seen previously (W04; Schmidt et
al. 2007).  This is likely a result of our strict quality cuts, which
removes the distant, late-type stars from the sample.  The more
distant stars are more likely to be older and therefore
inactive. However, the low signal-to-noise in those spectra makes
their activity state uncertain, and they are therefore excluded from
the analysis.

\begin{figure}
\plotone{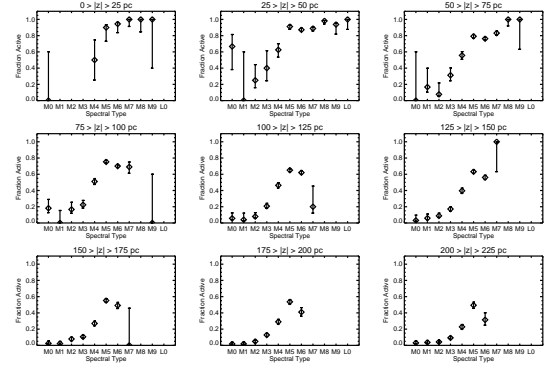}
\caption{The fraction of active stars as a function of spectral type
 for 9 different regions in vertical distance from the Plane.  By
  tracing individual spectral types over each distance bin, we see
  that distance plays an important role in determining the shape of
  Figure \ref{figure:actfrac}; most of the stars are active in the
  nearby bins and the activity fraction declines as we examine more
  distant stars.}
\label{figure:disteff} 
\end{figure}

Figure \ref{figure:actfrac} highlights an important point about 
the selection effects surrounding previous determinations of the
activity fraction as a function of spectral type.  Because the
activity fraction at a given spectral type changes as a function of
Galactic height (W04; W06), the sample selection will
bias the observed activity fraction.  We demonstrate this effect by
plotting the activity fraction as a function of spectral type and
constraining the absolute Galactic height of the stars to 9 different
regions (Figure \ref{figure:disteff}). It is clear that the volume
in which the data are observed plays a large role in the resulting activity fractions.

\begin{figure*}
\epsscale{1}
\plotone{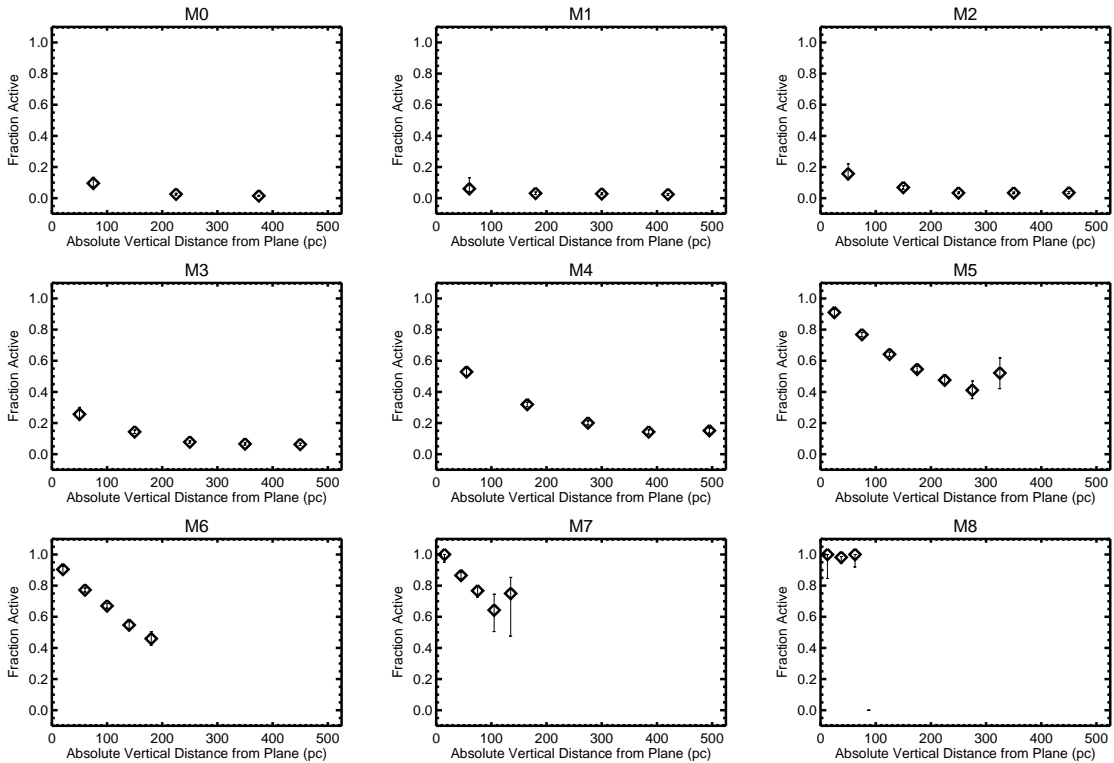}
\caption{The fraction of active stars as a function of absolute
  distance from the Galactic Plane for M0-M8 dwarfs.  The activity
  fraction decreases with Galactic height at all spectral types.  The
  magnitude and slope of the decline varies as a function of spectral
  type.  This effect is likely caused by the combination of thin disk
  dynamical heating and a rapid decrease of magnetic activity after a
  characteristic time.} 
\label{figure:allfracgrid} 
\end{figure*}

The distance effect can be clearly seen by expanding on the study of
W06 and examining how the activity fraction changes as
a function of vertical height for each spectral type. Figure
\ref{figure:allfracgrid} shows the activity fraction as a function of
absolute height above the Plane for spectral types M0-M8 (data have
been folded across the Plane to increase the signal).  There are
not enough stars in the later-type bins to justify their analysis.  In
almost every case, the stars closer to the Galactic Plane have larger activity
fractions than those further from the Plane.  The amplitudes and
slopes of the decrease are different for each spectral type.
Dynamical models with activity lifetimes are 
fit to these data and discussed in \S4.

\begin{figure}
\plotone{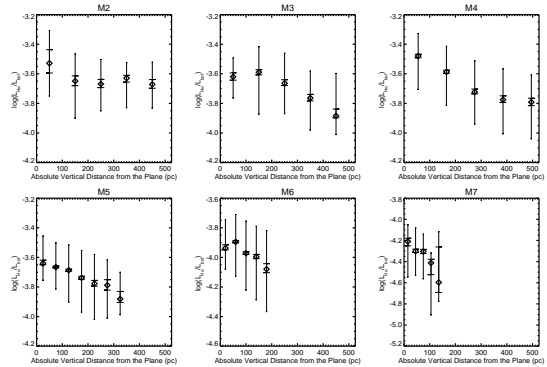}
\caption{Median L$_{\rm{H\alpha}}$/L$_{bol}$ as a function of absolute
  vertical distance from the Plane.  The narrow error bars indicate
  the spread of L$_{\rm{H\alpha}}$/L$_{bol}$ values while the wide
  error bars are the uncertainties in the median relations.  Although
  the spread in L$_{\rm{H\alpha}}$/L$_{bol}$ is large at each bin,
  there is a slight but significant decreasing trend with distance
  from the Galactic Plane.  The decrease is marginally consistent
  with a Skumanich (1972) type power law age-activity relation
  (t$^{-\alpha}$).  However, a unique value of $\alpha$ cannot explain
  all of the trends and the activity decrease is likely more complicated.}
\label{figure:lbolgrid} 
\end{figure}

As previously seen in M7 dwarfs (W06), most of the spectral types show
some evidence that the level of activity (quantified by
L$_{\rm{H\alpha}}$/L$_{bol}$; Hawley et al. 1996) also decreases as a
function of vertical height above the Plane.  Figure
\ref{figure:lbolgrid} shows the mean L$_{\rm{H\alpha}}$/L$_{bol}$
distribution for M2-M7 stars as a function of vertical distance from
the Plane.  The other types were excluded because of the small number
of active stars at early spectral types and small vertical extent at
late spectral types.  Although the spread of the distribution is large
(large error bars) in each bin, the uncertainty in the mean (small
error bars) is small, and indicates a small but significant decrease
in activity with distance.  This trend runs opposite to what one might
expect from a magnitude-limited selection effect; stars with more
activity should be easier to see further away.  This decrease is
marginally consistent with a Skumanich (1972) type power law
age-activity relation (t$^{-\alpha}$).  However, a unique value of
$\alpha$ cannot explain all of the trends, and the activity decrease
is likely more complicated.

In addition, the lack of stars with small L$_{\rm{H\alpha}}$/L$_{bol}$
at small distances from the Plane suggests that activity does not
continue a slow decline forever, but undergoes a rapid decrease at
some point during a star's lifetime.  There are stars in the DR5
sample that have H$\alpha$ EW $<$ 1\AA\ but do not get included in our
main study and have smaller L$_{\rm{H\alpha}}$/L$_{bol}$ values.
However, there is still a population of nearby, high signal-to-noise
stars in which we could detect H$\alpha$ activity if it were there -
but we don't.  Our ability to observe inactive stars is confirmed by
other studies including the high resolution spectroscopic study of
Rauscher \& Marcy (2006) and the Bochanski et al. (2007a) spectral
templates, where thousands of our inactive SDSS spectra were co-added
to produce extremely high signal-to-noise template spectra. The
inactive templates as well as many of the Rauscher \& Marcy (2006)
stars show no signs of H$\alpha$ activity, confirming our ability to
measure a star when it is in the inactive phase.

\subsection{Metallicity}
In addition to having less activity and greater dynamical heating,
stars formed earlier in the Galaxy's history should have fewer metals;
we should see a decrease in metallicity as a function of vertical
distance from the Plane.  Currently, there are no robust techniques
for translating low resolution spectra of M dwarfs to an absolute
metallicity scale.  Woolf and Wallerstein (2006) attempted to
calibrate M dwarf metallicities using the molecular indices described
in in previous studies (Reid, Hawley \& Gizis 1995; Gizis 1997), but
were thwarted by small sample size.  High resolution follow-up of
recent large surveys of cool, nearby stars that span a range of
metallicities (e.g. this sample; Marshall 2007), will allow for larger
datasets to properly calibrate the metallicity scale.  In the
meantime, we can use molecular indices to determine relative metal
content.  Recent studies have used a combination of the Gizis
(1997) indices (CaH2, CaH3 and TiO5) to identify low mass subdwarfs
and extreme subdwarfs (Lepine, Shara \& Rich 2003; Burgasser \&
Kirkpatrick 2006) .  We adapt the Lepine two-dimensional metallicity
space to a single ratio, (CaH2+CaH3)/TiO5, and use this as a proxy for
metallicity at a given spectral type (a similar technique was used in
Bochanski et al. 2007b).

\begin{figure*}
\plotone{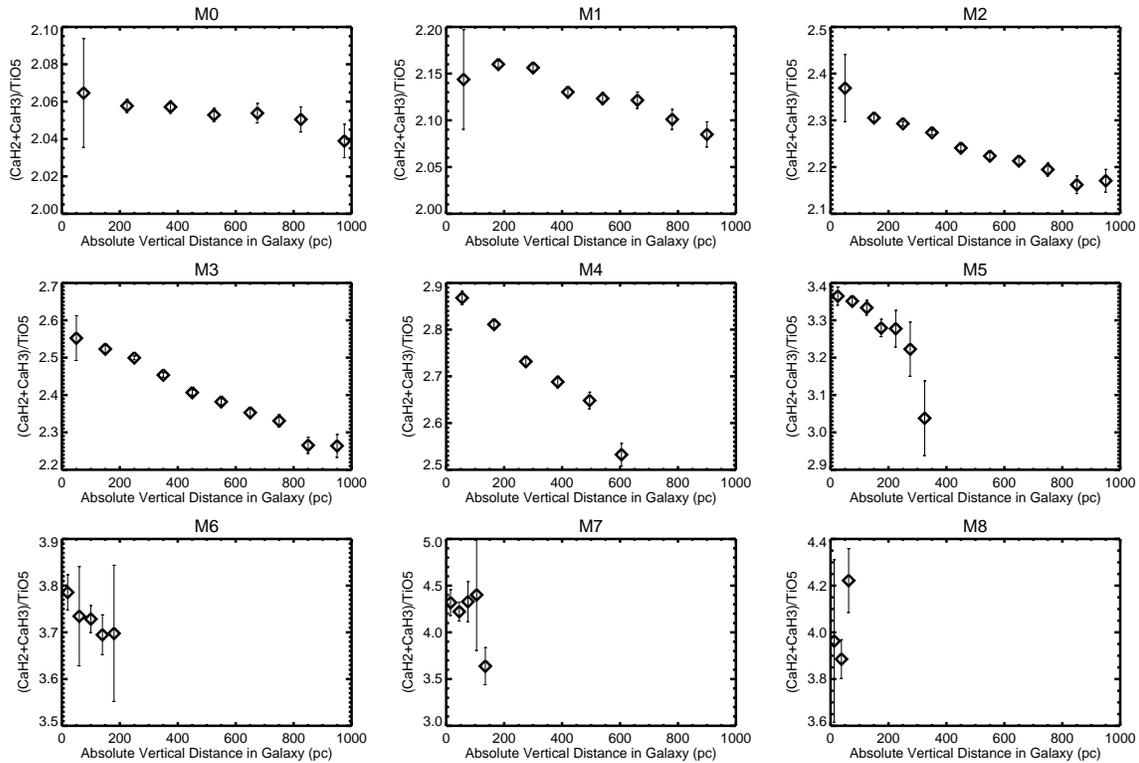}
\caption{The metal sensitive ratio (CaH2+CaH3)/TiO5 as a function of
  absolute vertical distance from the Plane for M0-M8 dwarfs.  Stars
  further from the Galactic Plane are more metal-poor.  Error bars
  represent the uncertainties in the mean values for each bin. Although we
  cannot yet tie this ratio to an absolute metallicity scale,
  our results demonstrate that future M dwarf studies should be able
  to determine the metallicity evolution of the Milky Way thin disk in
a statistically robust way.} 
\label{figure:met} 
\end{figure*}

Figure \ref{figure:met} shows the metal sensitive ratio
((CaH2+CaH3)/TiO5) as a function of vertical distance above the Plane
for M0-M8 dwarfs.  A decrease in metallicity as a function of Galactic
height is clearly evident, indicating that the more distant stars have
fewer metals.  For the latest-type dwarfs, the small number of usable
stars does not allow for an accurate analysis, but for the early-mid
types the metallicity trend is both pronounced and significant.  In
the future, when molecular indices can be reliably tied to an absolute
metallicity scale, we will be able to use each spectral type
independently to calibrate the metallicity evolution of the Milky Way
disk.

\section{Simulations}

We used the simple 1-D dynamical model described in W06 to trace the
vertical dynamics of stars as a function of time.  For each spectral
type, we assumed a constant star formation rate, and injected a new
population of 50 stars at the Galactic midplane every 200 Myr, for a
total simulation time of 10 Gyrs.  Each new group of stars began with
a randomly drawn velocity dispersion of 8 km s$^{-1}$ (Binney et
al. 2000) and had a new position and velocity computed every 0.1 Myr.
We used the ``leap-frog'' integration technique (Press et al. 1992)
and the vertical Galactic potential from Kuijken \& Gilmore (1989) and
Siebert et al. (2003) to track the vertical dynamics of each star in
the simulation.

We simulated dynamical heating by altering the velocities (energies)
of stars such that their new velocity dispersions would match
a $\sigma \propto t^{0.5}$ relation (Wielen 1977;
Fuchs et al. 2001; H\"anninen \& Flynn 2002).  The velocities were
also scaled by the square-root of the mass appropriate for each
type (as defined by Reid \& Hawley 2005).  We varied the heating
energy about this mean value and added randomly drawn velocities
when stars were within a given distance from the Plane.  This
distance (or ``region of influence''; see W06) is a way to
parameterize the cross section of interactions with molecular clouds
during a vertical Plane crossing and it is modeled to be symmetric about
the Galactic midplane.  In a given orbit,
a star may have no interactions with molecular clouds, or it may run
into the full extent of a molecular region.  Thus, to first order,
``the regions of influence'' can be thought of as the product of the
scale height of molecular gas and the fraction of Plane crossings that
result in a dynamical kick.  The size of the ``region of influence'' is varied from $\pm$0.5 pc to $\pm$5 pc in
intervals of 0.5 pc.  In total we ran 70 simulations for each M dwarf
spectral type.  Our simulations tracked the velocity, position and age
of each star.

\begin{figure}
\epsscale{1}
\plotone{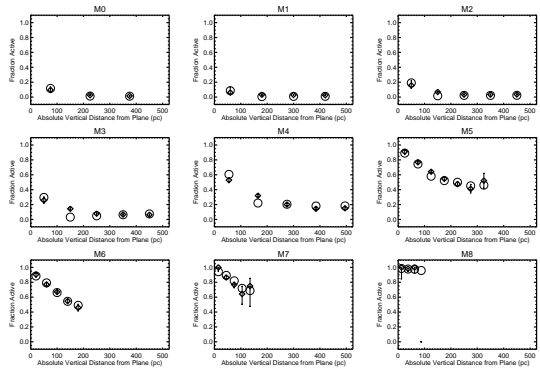}
\caption{Active fraction as a function of absolute vertical distance
  from the Plane (diamonds) compared with the best-fit simulation
  data (open circles).  All simulated data uses a ``region of
  influence'' of 1 pc.  The simple 1-D simulations fit the observed
  trends well at all spectral types.} 
\label{figure:allfracgridmod} 
\end{figure}

After the dynamical simulations were complete, we introduced a
timescale for magnetic activity to the simulated data.  Specifically,
we assigned an activity lifetime to every star of a given spectral
type.  We varied these lifetimes from 0 Gyr to 9 Gyr in 0.1 Gyr
intervals for each individual simulation. From the resulting data, we
can derive activity fractions as a function of vertical distance from
the Plane and examine how the simulated properties compare to the
observed data.

We compared the simulated activity fractions to the observed activity
fraction as a function of galactic height and performed a chi-squared minimization, fitting for
``region of influence'' size, dynamical heating energy and activity
lifetime.  Because the ``region of influence'' should be the same for
all spectral types, we first solved for the best-fit value and found it to
be 1 pc, slightly smaller than the 3 pc found in W06.  Although 1 pc
is close to the edge of the explored grid (0.5 pc), it produces a
significantly better fit to the data as compared to the runs that use
0.5 or 1.5 pc values.  Then, using a fixed
``region of influence'' of 1 pc, the energies and
activity lifetimes were determined for each spectral type.  
Figure \ref{figure:allfracgridmod}
shows the observed activity fraction as a function of vertical
distance from the Plane with the simulated trends
overplotted for the best-fit simulations.

\begin{figure}
\epsscale{1}
\plotone{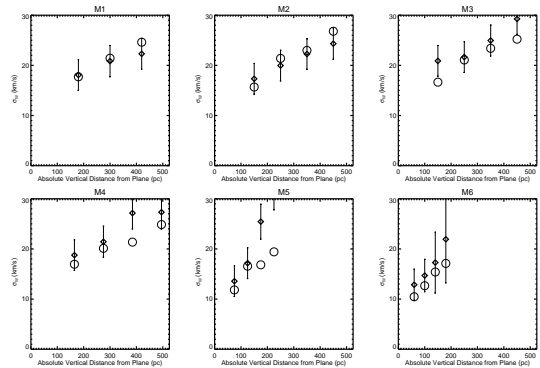}
\caption{Vertical velocity dispersions as a function of vertical
  distance from the Plane for the DR5 sample (diamonds) compared to
  the simulations (open circles).  Only spectral types
  that have large numbers of nearby ($<$ 500 pc) stars with measured
  proper motions are shown.  Our simple dynamical simulations are
  able to reproduce nearly all of the velocity data.}
\label{figure:allvelgridmod} 
\end{figure}

To ensure that our simulations were producing dynamically accurate
results, we compared the vertical velocity dispersions from our
proper motion matched sample to the simulated velocity
dispersions. The velocity dispersions were measured using the 
probability analysis described in Bochanski et al. 2007b (see
also Reid et al. 1995).   
Figure \ref{figure:allvelgridmod} shows the velocity dispersions
as a function of absolute vertical distance from the Plane for both
the DR5 sample and the simulations.  For several
spectral types, there are a limited number of stars with good proper
motions.  Therefore, only the M1-M6 spectral types have been included.
The data and model velocities are in excellent agreement for most of
the measured velocity bins.  Deviations from the models are likely due
to the simplicity of our dynamical models, which do not allow for
such known complications as dynamical substructure.

\begin{deluxetable}{lccccc}
\tablewidth{80pt}
\tablecolumns{3} 
\tabletypesize{\small}
\tablecaption{Activity Lifetimes of M dwarfs}
\renewcommand{\arraystretch}{.6}
\tablehead{
\colhead{Spectral Type}&
\colhead{Age (Gyr)}}
\startdata
\vspace{1mm}
M0&$0.8\pm^{\ 0.6}_{\ 0.6}$\\
\vspace{1mm}
M1&$0.4\pm^{\ 0.4}_{\ 0.4}$\\
\vspace{1mm}
M2&$1.2\pm^{\ 0.4}_{\ 0.4}$\\
\vspace{1mm}
M3&$2.0\pm^{\ 0.5}_{\ 0.5}$\\
\vspace{1mm}
M4&$4.5\pm^{\ 0.5}_{\ 1.0}$\\
\vspace{1mm}
M5&$7.0\pm^{\ 0.5}_{\ 0.5}$\\
\vspace{1mm}
M6&$7.0\pm^{\ 0.5}_{\ 0.5}$\\
\vspace{1mm}
M7&$8.0\pm^{\ 0.5}_{\ 1.0}$\\
\enddata
\label{table:activeage}
\end{deluxetable}

\begin{figure*}
\plotone{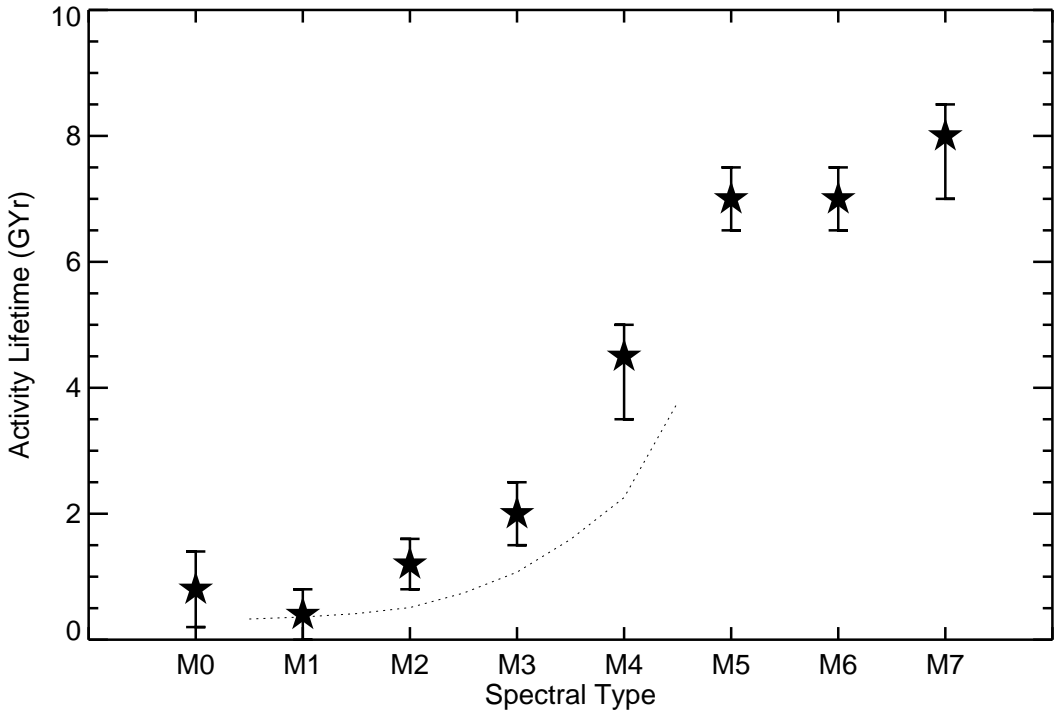}
\caption{The activity lifetimes of M dwarfs as determined by comparing
  the DR5 spectroscopic data to the 1-D simulations (stars).  The
  Hawley et al. (2000) age-activity relation is overplotted for
  comparison (dotted).  As predicted, the Hawley relation provides a lower
  limit to the ages.  We find that there is
  a significant increase in activity lifetimes between spectral types
  M3 to M5, possibly
  indicating a physical change in the production of magnetic fields
  as full convection sets in.  The uncertainties indicate the spread 
  of the model fits at each spectral type. } 
\label{figure:activeage} 
\end{figure*}

The final result of our analysis is an estimate of the activity
lifetime at each spectral type.   The values are given in 
Table \ref{table:activeage}
and shown in Figure \ref{figure:activeage}.
The error bars indicate the range of
activity lifetimes for all of the simulations at each spectral type.
The dotted line is the Hawley et al. (2000) activity-age relation.
The activity lifetimes increase to later spectral types and as predicted, our
results fall above the Hawley relation and
extend to later spectral types.  Due to the small sample size beyond a
spectral type of M7, we only include lifetimes for M0-M7 dwarfs.
There is a large jump in the activity lifetime at the fully convective boundary between types M3-M5, suggesting that a change in the magnetic dynamo mechanism at this boundary leads to much longer lasting magnetic activity in later type M dwarfs.  The derived trend also increases our understanding of
the selection effects responsible for the activity fractions shown in
Figures \ref{figure:actfrac} and \ref{figure:disteff}; later type
stars are seen to be more active because they are both closer
\emph{and} their active lifetimes are longer.

\section{Summary}

Using over 44,000 spectroscopically confirmed M and L dwarfs, we have
expanded the analysis of W06 and completed a detailed study of the
magnetic activity, metallicity and dynamics of cool stars in the
SDSS. The results of our analysis are summarized below.

1. We have compiled the largest spectroscopic sample of low mass stars
ever assembled.  By matching our spectra to SDSS and 2MASS photometry,
we compute the mean optical/IR colors for each spectral type with
unprecedented robustness.  Although these colors represent the best
determination of optical/IR colors for cool stars to date, they may
still be affected by the non-uniform sample selection.

2. We show that activity fractions decrease as a function of absolute
height above the Galactic Plane for all M dwarf spectral types and
that the magnitude and
slope of this effect is different for each spectral type.

3. The amount of activity as quantified by
L$_{\rm{H}\alpha}$/L$_{bol}$ decreases slightly as a function of
vertical distance from the Plane.  However,
L$_{\rm{H}\alpha}$/L$_{bol}$ does not continue to decrease to the
smallest detectable values, providing evidence for a rapid decrease of
magnetic activity after some characteristic lifetime.

4. The metallicity sensitive ratio (CaH2+CaH3)/TiO5 also decreases
with increased distance from the Plane.  Although we cannot yet calibrate
this ratio to an absolute metallicity scale, we suggest that future
studies will be able to use data of this type to determine the
metallicity evolution of the Milky Way using low mass dwarfs.

5. We use a 1-D dynamical simulation to create a realistic model
of M dwarf motions and positions over the lifetime of the Galaxy and
show that our models match the observed activity and velocity trends.

6. Comparing our activity data to the dynamical simulations, we
derive an age-activity relation for M dwarfs for spectral types
M0-M7. We find that there is a strong increase in the active lifetimes
of stars in the spectral type range from M3-M5, which may be due to the
onset of full convection and subsequent change in the mechanism for
producing and storing magnetic fields in the stellar interior.

This last point is the main result of the paper and deserved further
elucidation.  The sharp increase in activity lifetime after spectral
type M3 (where the stars are predicted to become fully convective) may
be an empirical indication of the change from a solar-like,
rotationally-dependent magnetic dynamo in the earlier type stars, to a
turbulent dynamo (that may or may not be dependent on rotation) in the
later type stars.  The short activity lifetimes of the early type M
dwarfs would then be naturally explained by an extension of solar type
rotation-activity relations (e.g. Skumanich) to these stars, and the
timescale would be set by their spindown time. Although previous
studies have claimed evidence for a rotation-activity relation
extending past the M3 convective transition and into the brown dwarf
regime (Delfosse et al. 1998; Mohanty \& Basri 2003), these studies
may be biased by samples of nearby, young, active M dwarfs.  Future rotation
studies of inactive late-type M dwarfs are needed to confirm these
results.

The long activity lifetimes we have found for the fully convective
stars, where the magnetic fields are presumably generated by a
turbulent dynamo mechanism, set significant constraints on the
operation of the dynamo and the emergence of the surface fields to
power the activity. Future models must successfully predict long
(several Gyr) lifetimes, and the lifetimes must be mass dependent
(increasing with lower mass).

Our age-activity relation is consistent with previous findings for
early type M dwarfs and we extend the relation to stars near the M dwarf/brown
dwarf frontier.  The faintness of M dwarfs in old star clusters limits our
ability to observe this relation directly. However, future
spectroscopic studies should be able to test our age-activity relation with data for clusters of known (older) ages.

Many other studies are possible with this large spectroscopic sample.
Future papers will focus on the detailed dynamics of M dwarfs as probes
of Galactic structure (Hawley et al. 2008; West et al. 2008),
investigate the flare rate of active stars (Hilton et al. 2008) and
extend our DR5 sample to the lower mass L dwarfs (Schmidt
et al. 2008).  

The advent of large surveys such as SDSS and 2MASS has allowed for
many statistically robust studies of M dwarfs, the most numerous yet
faintest stellar constituents of the Milky Way.  Future large surveys
such as Pan-STARRS and LSST will add much to photometric studies of
low-mass stars but will not have spectroscopic components.  Therefore,
this SDSS low-mass star spectroscopic sample should remain useful
for the foreseeable future.

\section{Acknowledgments}
AAW acknowledges the support of NSF grant 0540567. SLH and JJB are
supported by NSF grant AST02-05875 and AST067644.  KRC is supported by
a Spitzer Postdoctoral Fellowship. The authors would like to thank
Matthew Browning, Lucianne Walkowicz, Tom Quinn, Adam Burgasser, and
Gibor Basri for useful discussions while completing this project.

    Funding for the SDSS and SDSS-II has been provided by the Alfred P. Sloan Foundation, the Participating Institutions, the National Science Foundation, the U.S. Department of Energy, the National Aeronautics and Space Administration, the Japanese Monbukagakusho, the Max Planck Society, and the Higher Education Funding Council for England. The SDSS Web Site is http://www.sdss.org/.

    The SDSS is managed by the Astrophysical Research Consortium for the Participating Institutions. The Participating Institutions are the American Museum of Natural History, Astrophysical Institute Potsdam, University of Basel, Cambridge University, Case Western Reserve University, University of Chicago, Drexel University, Fermilab, the Institute for Advanced Study, the Japan Participation Group, Johns Hopkins University, the Joint Institute for Nuclear Astrophysics, the Kavli Institute for Particle Astrophysics and Cosmology, the Korean Scientist Group, the Chinese Academy of Sciences (LAMOST), Los Alamos National Laboratory, the Max-Planck-Institute for Astronomy (MPIA), the Max-Planck-Institute for Astrophysics (MPA), New Mexico State University, Ohio State University, University of Pittsburgh, University of Portsmouth, Princeton University, the United States Naval Observatory, and the University of Washington.

\bibliographystyle{apj}







\end{document}